%% file: multitap_article.tex
\documentclass{IEEEtran}

\usepackage{amsmath}
\usepackage{amssymb}
\usepackage{amsthm}
\usepackage{tikz} 
\usepackage{pgfplots}
\usepackage{algorithmic}
\usepackage{algorithm}
\usepackage{bm}

\theoremstyle{definition}

\usepackage{cite}

\usetikzlibrary{plotmarks}
\pgfplotsset{compat=newest} 
\pgfplotsset{plot coordinates/math parser=false}
\newlength\figurewidth
\title{Multi-tap Digital Canceller for Full-Duplex Applications}
\author{\IEEEauthorblockN{Paul Ferrand,~\IEEEmembership{Member,~IEEE} and Melissa Duarte,~\IEEEmembership{Member,~IEEE}} \thanks{The authors are with the Mathematical and Algorithmic Sciences Laboratory, Huawei Technologies France, 92100 Boulogne-Billancourt, France. Emails: paul.ferrand@huawei.com and melissa.duarte@huawei.com. \newline 978-1-5090-3009-5/17/\$31.00 \textcopyright 2017 IEEE.}}

\definecolor{col1}{rgb}{0.0000,0.4470,0.7410}%
\definecolor{col2}{rgb}{0.8500,0.3250,0.0980}%
\definecolor{col3}{rgb}{0.9290,0.6940,0.1250}%
\definecolor{col4}{rgb}{0.4940,0.1840,0.5560}%
\definecolor{col5}{rgb}{0.4660,0.6740,0.1880}%
\definecolor{col6}{rgb}{0.3010,0.7450,0.9330}%

\begin{document}


\maketitle
\thispagestyle{empty}

%

\begin{abstract}
\input{abstract}
\end{abstract}

\input{intro}

\input{singletap}
\input{multitap}

\input{conclusion}
\input{appendix}
\label{sec:results}

\bibliographystyle{IEEEtran}
\bibliography{IEEEabrv,multitap_article}

\end{document}

%% file: abstract.tex
We identify phase noise as a bottleneck for the performance of digital self-interference cancellers that utilize a single auxiliary receiver---single-tap digital cancellers---and operate in multipath propagation environments.
Our analysis demonstrates that the degradation due to phase noise is caused by a mismatch between the analog delay of the auxiliary receiver and the different delays of the multipath components of the self-interference signal.
We propose a novel multi-tap digital self-interference canceller architecture that is based on multiple auxiliary receivers and a customized Normalized-Least-Mean-Squared (NLMS) filtering for self-interference regeneration.
Our simulation results demonstrate that our proposed architecture is more robust to phase noise impairments and can in some cases achieve 10~dB larger self-interference cancellation than the single-tap architecture.

%% file: intro.tex

\section{Introduction}
\label{sec:intro}

Since wireless radios share a common propagation medium for transmitting and receiving data, they were up to now limited to a half-duplex mode of operation.
In full-duplex mode, the proximity of the transmit and receive antenna generates a high powered self-interference signal at the receiver side, which shadows the intended signal from a remote transmitter.
Therefore, in a single time-frequency resource, the radio could either receive or transmit data, but not simultaneously.
This separation is inefficient since it basically halves the performance of the system as a whole.
A full-duplex wireless radio on the other hand can recover this performance loss, provided it would be able to handle the self-interference signal.
Recent advances in transceiver design have enabled this application at practical costs, and managed to restrict the self-interference below the noise floor at the receiver side \cite{Duarte2012,Korpi2016}.

However, it has been shown that self-interference cancellers are very sensitive to imperfections in the transceiver components \cite{Riihonen2012,Sahai2013,Ahmed2015a}.
Different architectures to deal with such imperfections have been proposed, each with its own advantages and limitations.
One recently proposed architecture for digital self-interference cancellation utilizes an auxiliary receiver that \emph{taps} a copy of the analog self-interference signal before the transmitter antenna.
This single-tap digital canceller architecture has been described in \cite{Ahmed2015a,Ahmed2015b,Choi2015,Korpi2014a,Korpi2014b,Li2011}.
One key advantage of this solution is that no estimation nor compensation of transmitter-side impairements is required; they basically have no effect in the performance of the system since the auxiliary receiver \emph{knows} the impaired self-interference signal and does not try to decode it \cite{Ahmed2015a}.
However, the receiver-side impairments can still degrade the performance.

In this paper, we identify that the performance of the auxiliary receiver architecture of \cite{Ahmed2015a,Ahmed2015b,Choi2015,Korpi2014a,Korpi2014b,Li2011} degrades in the presence of phase noise as a function of the delay spread of the channel---thereby confirming the observations of \cite{Ahmed2015a} where channels with shorter coherence bandwidth exhibited worse performance.
In contrast to \cite{Ahmed2015a}, our results stem from a time domain analysis that applies to single or multicarrier systems, while the results in \cite{Ahmed2015a} focused on the analysis of a multicarrier scenario.
We show that performance degradation due phase noise is fundamentally a problem of delay between the auxiliary receiver channel and the self-interference channel.
We provide an analysis of the degradation induced by the phase noise process; these results show similarities to the effect of oscillator phase noise on analog full-duplex cancellers \cite{Sahai2013}.
We then introduce in Section~\ref{sec:canceller} a new architecture that uses possibly many analog reference taps in the auxiliary receiver.
We use these multiples \emph{views} on the self-interference signal to design simple Normalized-Least-Mean-Squared (NLMS) adaptive filters, which automatically adapt to spread channels without having to estimate the phase noise process. 
These filters are reminiscent of approaches used in echo cancellation when the acoustic channel has very large delays \cite{Sharma2003}.
As simulation results will demonstrate, our proposed multi-tap canceller can achieve much larger self-interference cancellation than the single-tap one and is more robust to increasing delay spread.

%% file: singletap.tex

\section{Phase noise effect on the single-tap digital canceller}
\label{sec:pn_effect}
\begin{figure}
	\centering
	\includegraphics[width=\columnwidth]{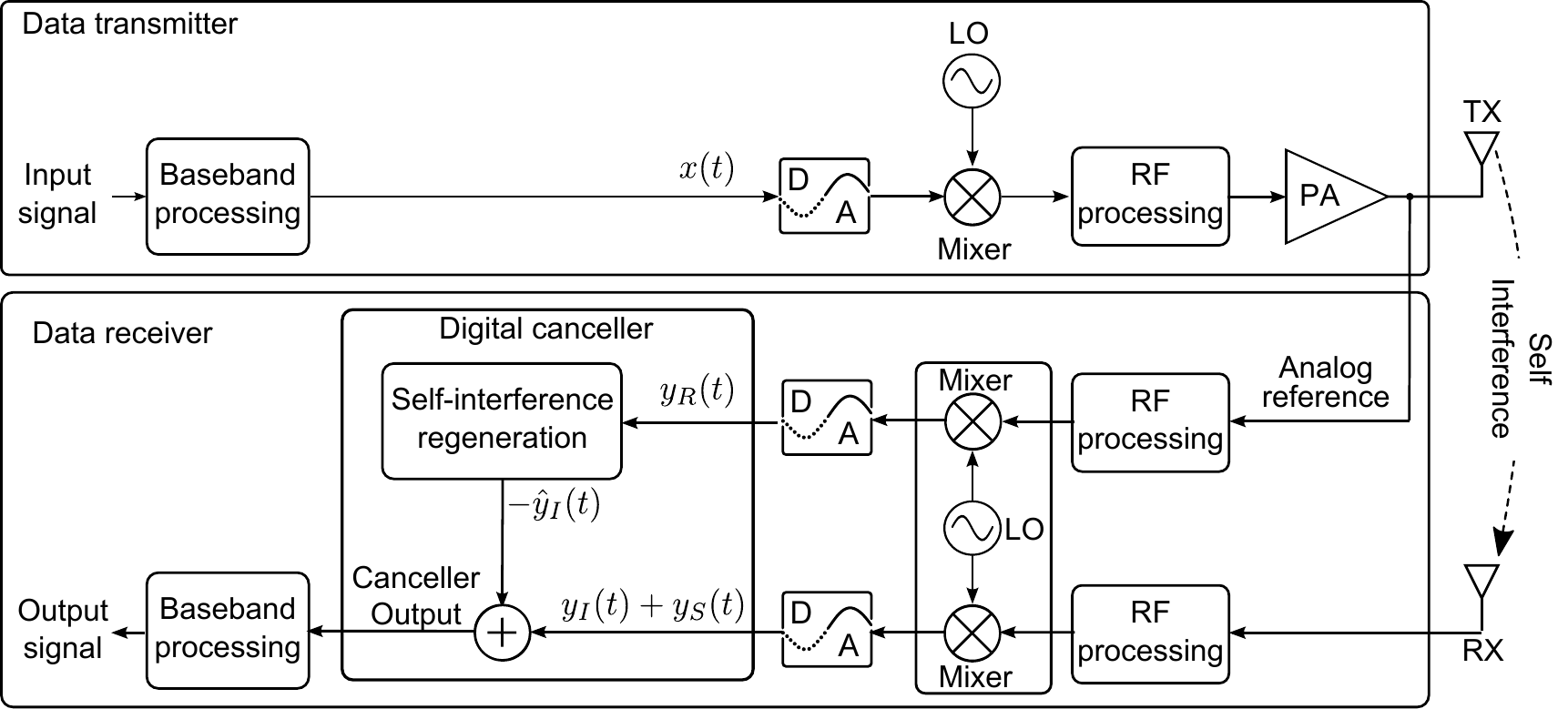}
	\caption{Full-duplex transceiver using a single-tap digital self-interference canceller.}
	\label{fig:singletap_bd}
\end{figure}
In this section we analyze the phase noise effect on the full-duplex transceiver shown in Fig.\ref{fig:singletap_bd}.
It implements time domain cancellation as in \cite{Korpi2014a,Korpi2014b} through a single analog reference tapped from the transmit signal.
The architecture consists in one transmitter chain with a digital to analog converter (D/A), a mixer fed by a local oscillator (LO) and RF processing that includes conventional filtering stages followed by a power amplifier (PA) whose output signal feeds the transmitter antenna.
On the receiver side, there is one receiver chain connected to the receiver antenna and one receiver chain connected to the tapped analog reference.
Each receiver chain consists of an analog to digital converter (A/D), a mixer fed by the LO (which is shared by all receiver mixers as in \cite{Korpi2014a,Ahmed2015a}) and RF procesing which includes filtering stages and a low noise amplifier (LNA)---which may not be required by the auxiliary receiver \cite{Korpi2014a}.
We assume there is no analog cancellation as in \cite{Ahmed2015a}; receiver saturation is avoided via the use of passive techniques \cite{Everett2014}.
This architecture is thus more suited for scenarios with either low transmit power or high passive suppression.
At the receiver, the signals at the output of the A/D converters are input to the time domain digital self-interference cancellation, which consists in subtracting an estimate of the self-interference signal from the signal coming from the receiver antenna.
In this paper we consider a self-interference regeneration that is implemented via an adaptive filtering stage.
This approach of using an adaptive filter has been briefly mentioned in \cite{Korpi2014b} but its performance was not evaluated. 
Note that we assume a calibration period where the equalizer is trained without the presence of the signal of interest $y_{S}(t)$.
A calibration period for estimation of the self-interference signal has been used in related work~\cite{Duarte2012,Korpi2014a}, where the authors have shown that it benefits the overall system in terms of net interference cancellation.
The penalty of this calibration period is obviously a reduction in the rate~\cite{Korpi2014b}.

We now analyze the residual self-interference for the full-duplex transceiver in Fig.\ref{fig:singletap_bd}, with an emphasis on the effect of the phase noise at the receiver and the relative delay between the reference and the self-interference.
Let the self-interference channel be a single path channel with attenuation $h_{I}$ and delay $\Delta_{I}$.
The received self-interference signal, including the effect of receiver phase noise $\phi(t)$, is equal to
\begin{equation}\label{eq:yI}
y_{I}(t) = h_{I}e^{\jmath\phi(t)}x(t-\Delta_{I}).
\end{equation}
Let $h_R$ and $\Delta_{R}$ denote the attenuation and delay of the analog reference signal path.
The reference signal at the receiver is equal to 
\begin{equation}\label{eq:yR}
y_{R}(t) = h_{R}e^{\jmath\phi(t)}x(t-\Delta_{R}).
\end{equation}
Let us define $h_{IR}=h_{I}/h_{R}$ and $\Delta_{IR} = \Delta_{I}-\Delta_{R}$.
We can write the received self-interference as a function of the reference signal as follows
\begin{equation}
\label{eq:eq_channel}
y_{I}(t) = h_{IR}e^{\jmath \left( \phi(t) - \phi(t-\Delta_{IR}) \right) } y_{R}(t-\Delta_{IR}).
\end{equation}
From the above equation we see that, due to the phase noise term $e^{\jmath \left( \phi(t) - \phi(t-\Delta_{IR}) \right) }$, the relation between $y_{I}(t)$ and $y_{R}(t)$ changes with each time sample.
This fast variation is hard to track for the adaptive filter that performs the self-interference regeneration.
Adaptive algorithms used in equalization will converge over time to the \emph{average} channel between the input and output. 
The average channel depends on the stochastic process $\phi(t) - \phi(t-\Delta_{IR})$.
Most commonly, $\phi(t)$ is modeled as Gaussian processes---e.g. Wiener processes or Ornstein-Uhlenbeck processes \cite{Ahmed2015a,Herbert2014}.
The distribution of $\phi(t)$ for all $t$ is then a Gaussian random variable with some variance that may depend on time~\cite{Steele2001}.
The difference process is thus also a Gaussian process, and usually has zero expectation; under mild conditions there is no trend in the mean and $\mathbb E[\phi(t)] = \mathbb E[\phi(t - \Delta_{IR})]$.
Assume then that the difference process is distributed as $\theta(t) = \phi(t) - \phi(t-\Delta_{IR}) \sim \mathcal N(0, \sigma_\epsilon^2(t))$.
Through the characteristic function of a Gaussian random variable, we have that
\begin{equation}\label{eq:weight_k}
 	\mathbb E\left[e^{\jmath \theta(t) }\right] = \exp\left( -\frac{\sigma_\epsilon^2(t)}{2}\right) = K(t).
\end{equation}
The reconstructed interference will then be
\begin{equation}\label{eq:yest}
\widehat{y}_{I}(t) = h_{IR}K(t)y_{R}(t-\Delta_{IR}).
\end{equation}

Using $\widehat{y}_{I}(t)$ in Eq.~(\ref{eq:yest}) to cancel the self-interference $y_{I}(t)$ will leave some residual interference after cancellation which is computed as 
\begin{equation}\label{eq:rinit}
r(t) = \left(y_{I}(t) + w_{I}(t)\right) - \left(\widehat{y}_{I}(t) + w_{R}(t)\right), 
\end{equation}
where $w_{I}(t)$ represents AWGN in the receiver of the self-interference signal and $w_{R}(t)$ represents residual error of the self-interference regeneration that is due to AWGN in the reference and self-interference signals.
By replacing (\ref{eq:yest}) and (\ref{eq:yR}) in (\ref{eq:rinit}) and doing some basic simplifications we obtain
\begin{align}
r(t) =& \left( e^{\jmath\phi(t)}-K(t)e^{\jmath\phi(t-\Delta_{IR})} \right) h_{I}x(t-\Delta_{I}) \nonumber \\
&+ w_{I}(t) - w_{R}(t),\label{eq:r}
\end{align}
The power of the residual self-interference is once again dependent on the statistics of the stochastic process $\theta(t)$.
Let $N_I = \mathbb E \left[ \left| w_{I}(t) \right|^2 \right]$ and $N_R = \mathbb E \left[ \left| w_{R}(t) \right|^2 \right]$ be the noise powers, and let $N = N_I + N_R$.
Furthermore, let $ P_I = \left|h_I\right|^2 \mathbb E \left[ \left| x(t) \right|^2 \right]$ be the self-interference power.
We can write the residual self-interference power as
\begin{align}
E \left[ \left| r(t) \right|^2 \right] &= P_I \mathbb E\left[\left|  e^{\jmath\phi(t)}-K(t)e^{\jmath\phi(t-\Delta_{IR})}\right|^2\right] + N \notag \\
&= P_I \left( 1 + K(t)^2 \right. \notag \\ 
& \left. - K(t) \left(\mathbb E\left[e^{\jmath \theta(t)}\right] + \mathbb E\left[e^{-\jmath \theta(t)}\right]\right)\right) + N \notag \\
&= P_I(1-K(t)^2) + N \label{eq:res}
\end{align}
Notice that the residual interference due to phase noise reduces to zero when $\Delta_{I} = \Delta_{R}$ for which $K(t)^2 = 1$ regardless of the phase process; the detrimental effect of phase noise only manifests itself when  $\Delta_{I} \neq \Delta_{R}$.
This observation implies that if the self-interference channel consists of only one path, then the delay $\Delta_{R}$ of the auxiliary receiver should be set as close as possible to the delay of the self-interference path.
However, achieving this is not possible when the self-interference channel is a multipath channel.
In such scenario it is not possible to match all the path delays by using a single auxiliary receiver: there will be a residual self-interference that stems from multipath components for which $\Delta_{I} \neq \Delta_{R}$.

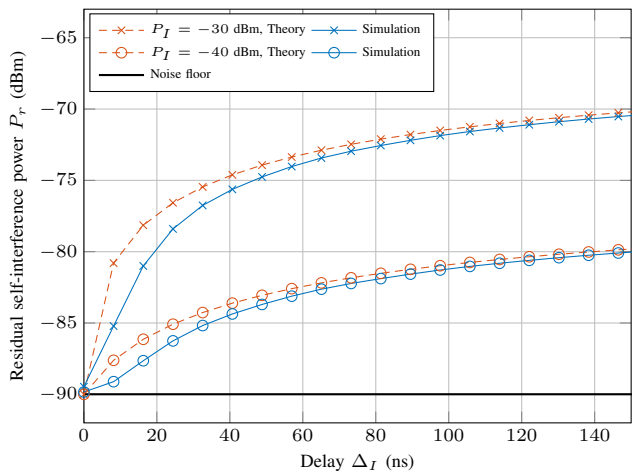
\begin{figure}[t!]
		\scriptsize
		\begin{tikzpicture}
		\begin{axis}[
			width=\figurewidth,
			height=0.8\figurewidth,
			xmajorgrids,
			ymajorgrids,
			xlabel=Delay $\Delta_I$ (ns),
			ylabel=Residual self-interference power $P_r$ (dBm),
			legend entries={
				{$P_I = -30$ dBm, Theory},
				{Simulation},
				{$P_I = -40$ dBm, Theory},
				{Simulation},
				{Noise floor},
			},
			legend columns=2,
			mark options={solid},
			xmin=0,
			xmax=150,
			ymin=-92,
			ymax=-63,
			legend style={at={(0.01, 0.99)}, anchor=north west, font=\tiny},
			legend cell align=left,
		]
			\addplot+[color=col2, densely dashed, mark=x] table[x index=0, y index=1] {spawc_fig_sp.data};
			\addplot+[color=col1, solid, mark=x] table[x index=0, y index=2] {spawc_fig_sp.data};
			\addplot+[color=col2, densely dashed, mark=o] table[x index=0, y index=3] {spawc_fig_sp.data};
			\addplot+[color=col1, solid, mark=o] table[x index=0, y index=4] {spawc_fig_sp.data};
			\addplot+[color=black, solid, thick, mark=none] table[x index=0, y index=7] {spawc_fig_sp.data};
		\end{axis}
		\end{tikzpicture}
	\caption{Residual interference power achieved by the single-tap digital self-interference canceller for the scenario of a single path channel with a varying path delay $\Delta_I$.
	}
	\label{fig:singletap_singlepath}
\end{figure}

\begin{figure}[b!]
	\scriptsize
	\begin{tikzpicture}
	\begin{axis}[
		width=\figurewidth,
		height=0.8\figurewidth,
		xmajorgrids,
		ymajorgrids,
		xlabel=Delay spread $|\Delta^{(2)}_{I} - \Delta^{(1)}_{I}|$ (ns),
		ylabel=Residual self-interference power $P_r$ (dBm),
		legend entries={
			{$\Delta_R = 0$ ns},
			{$\Delta_R = \delta_S/2$ ns},
			{$\Delta_R = \Delta^{(2)}_I$ ns},
			{Noise floor},
		},
		mark options={solid},
		xmin=0,
		xmax=150,
		ymin=-92,
		ymax=-63,
		legend style={at={(0.01, 0.99)}, anchor=north west, font=\tiny},
		legend cell align=left,
	]
		\addplot+[color=col1, solid, mark=square] table[x index=0, y index=1] {spawc_fig_2p.data};
		\addplot+[color=col1, solid, mark=o] table[x index=0, y index=2] {spawc_fig_2p.data};
		\addplot+[color=col1, solid, mark=x] table[x index=0, y index=3] {spawc_fig_2p.data};
		\addplot+[color=black, solid, thick, mark=none] table[x index=0, y index=7] {spawc_fig_2p.data};
	\end{axis}
	\end{tikzpicture}
	\caption{ Residual interference power achieved by the single-tap digital self-interference canceller for the scenario of a 2-path channel with with absolute delay spread $|\Delta^{(2)}_{I} - \Delta^{(1)}_{I}|$.
	}
	\label{fig:singletap_multipath}
\end{figure}
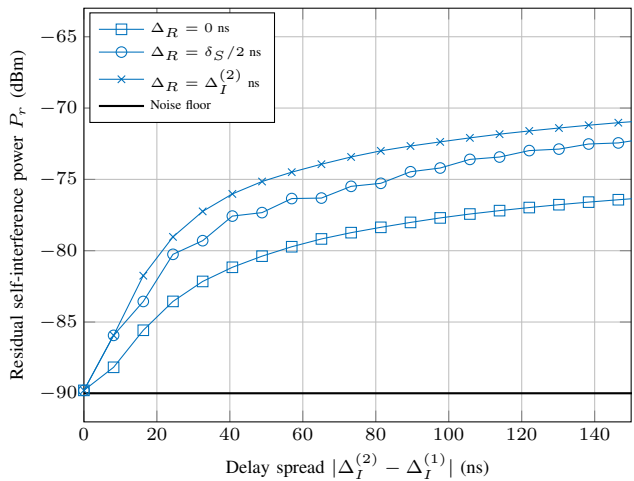

The above observations are verified in Fig.~\ref{fig:singletap_singlepath}.
The simulation results are obtained from a simulation of the transceiver in Fig.~\ref{fig:singletap_bd} and an OFDM signal $x(t)$.
The details of the simulation parameters are presented in the appendix.
Fig.~\ref{fig:singletap_singlepath} shows results for the case of a single path self interference channel and $\Delta_R = 0$.
We observe the residual self-interference power $P_r$ increases as $\Delta_{I}-\Delta_{R}$ increases and results verify the closed form expression derived in (\ref{eq:res}). The difference between simulations and our theoretical result can stem from the fact that, for simplicity, our analysis did not include the effect of pulse shaping and upsampling at the transmitter and matched filtering and  downsampling at the receiver. We did include these effects in the simulator, as explained in the appendix.

Fig.~\ref{fig:singletap_multipath} shows results for the case of a channel with two paths, where the first past has path loss $10\log_{10}|h^{(1)}_{I}|^2=-50$~dB and a delay $\Delta^{(1)}_{I}=0$~ns and the second path has a path loss of $10\log_{10}|h^{(2)}_{I}|^2=-60$~dB and delay $\Delta^{(2)}_{I}$ that varies between 0 and 150~ns, this choice of values is based on the time response channel measurements reported in \cite{Everett2014}.
We show results for different values of the reference signal delay~$\Delta_{R}$.
The results in Fig.~\ref{fig:singletap_multipath} show that the performance of the digital canceller degrades---and the residual power increases---as the delay spread $\delta_S = \Delta^{(2)}_{I} - \Delta^{(1)}_{I}$ for this multipath channel increases.
In other words, in the presence of phase noise, the performance of the digital canceller degrades as the delay spread of the channel increases and this is true for any value of $\Delta_{R}$.
The choice of $\Delta_R=0$ is what works best because this matches the delay of the strongest self-interference path.
The choice of $\Delta_R=\Delta^{(2)}_I$ matches the delay of the weakest path, thus the strongest path is not properly matched and results in higher residual.
The choice of $\Delta_R=\delta_S/2$ matches better the strongest path hence it is better than when $\Delta_R=\Delta^{(2)}_I$.

%% file: multitap.tex

\section{The multi-tap canceller}
\label{sec:canceller}

\begin{figure}[b!]
	\centering
	\includegraphics[width=\columnwidth]{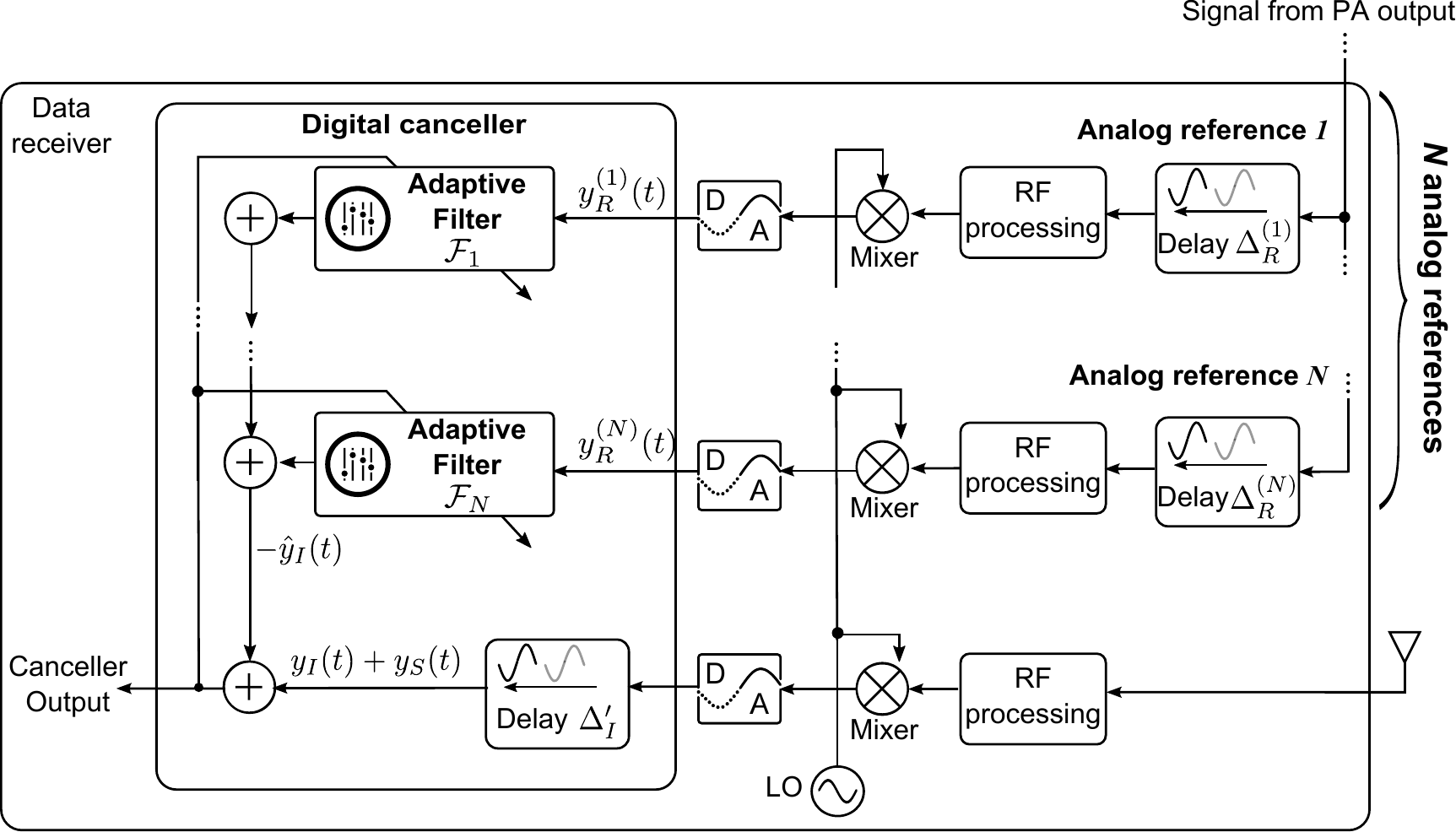}
	\caption{Proposed multi-tap digital self-interference canceller.}
	\label{fig:multitap_bd}
\end{figure}

Overall, using a single tap can effectively eliminate the residual due to phase noise whenever the self-interference channel is a single path channel and this can be achieved by setting $\Delta_{R} = \Delta_{I}$.
However, due to phase noise at the receiver, the single tap architecture will not be able to reduce the self-interference to the noise floor when the self-interference channel consists on multiple paths.
The key insight in our improved self-interference canceller is that, ideally, if we were to have 2 reference signals with different delays such that $\Delta^{(1)}_R = \Delta^{(1)}_I$ and $\Delta^{(2)}_R = \Delta^{(2)}_I$, then we could choose to match each path in the channel with the corresponding reference and avoid the degradation due to phase noise.

\subsection{Self-interference regeneration using multiple references}
\label{subsec:canceller_intuition}
We generalize this architecture in Fig.~\ref{fig:multitap_bd} and propose the use of $N$ reference signal branches.
Note that the delays of each branch need to be radio-frequency components, since they have to occur before the downconversion of the signal.
Note also that the branch connected to reference signal $n$, with delay  $\Delta^{(n)}_{R}$, can only be used to regenerate self-interference components that arrive with a delay larger than $\Delta^{(n)}_{R}$.
Consequently we apply a digital delay $\Delta'_{I} \geq \Delta^{(N)}_{R}$ to the received self-interference signal to guarantee that all the filter branches are useful.
Using $h^{(n)}_R$ and $\Delta^{(n)}_{R}$ to denote the attenuation and delay of the $n$-th analog reference signal, we write this signal as
\begin{equation}\label{eq:yRn}
y^{(n)}_{R}(t) = h^{(n)}_{R}e^{\jmath\phi(t)}x(t-\Delta^{(n)}_{R}).
\end{equation}
Assume that the self-interference signal consists of $M$ multipath components, and let $t' = t - \Delta'_I$ be the relative time at the receiver.
The observed self-interference after the digital delay $\Delta'_{I}$ is given by
\begin{align}\label{eq:yIm}
	y_{I}(t) = \sum_{m=1}^{M} h^{(m)}_{I}e^{\jmath\phi(t')}x(t'-\Delta^{(m)}_I),
\end{align}
where $h^{(m)}_{I}$ and $\Delta^{(m)}_I$ denote respectively the attenuation and delay that the self-interference experiences over path $m$.
Let us define $h^{(m,n)}_{IR}=h^{(m)}_{I}/h^{(n)}_{R}$ and $\Delta^{(m,n)}_{IR} = \Delta^{(m)}_{I}-\Delta^{(n)}_{R}$.
The difference phase process is then $\theta^{(m,n)}(t') = \phi\left(t'\right) - \phi\left(t'-\Delta^{(m,n)}_{IR}\right)$.
In a generalization of \eqref{eq:eq_channel}, we can write the received self-interference using the $n$-th reference as follows
\begin{align}\label{eq:yImn}
	y_{I}(t) = \sum_{m=1}^{M} h^{(m,n)}_{IR}e^{\jmath \theta^{(m,n)}(t')}y^{(n)}_{R}(t'-\Delta^{(m,n)}_{IR}).
\end{align}
As before, the farthest the delay of the reference path $\Delta^{(n)}_{R}$ from the delay of the multipath component $\Delta^{(m)}_{I}$, the more degraded the reconstruction will be in average.
A direct solution to would be to average the reconstructed signals from the various references, in an effort to reduce the variance of the noise in the reconstruction~\cite{Sharma2003}.
However, tradeoffs have to be made between the adaptive filter lengths and the number of references as to avoid noise enhancement effects.
An alternative solution is to ensure that only the references whose $\Delta^{(n)}_{R}$ is closest to the delay $ \Delta^{(m)}_{I}$ of the $m$-th path are chosen to reconstruct this specific path.
We now detail this second solution.

\subsection{An example of self-interference regeneration based on NLMS and using multiple references}
\label{subsec:canceller_LMS}

Consider the filter $\mathcal{F}_{n}$ with $L$ taps.
This filter has as an input the analog reference with delay $\Delta^{(n)}_{R}$.
Let $T_s$ be the sample period of the input to the filter and thus the time separation between the filter taps.
Due to the digital delay $\Delta'_{I}$ the filter $\mathcal{F}_{n}$ will put the largest weight to tap $\lceil \Delta'_{I} / T_s \rceil$ when regenerating a self-interference signal with $\Delta^{(m)}_{I} = \Delta^{(n)}_{R}$.
In other words, tap number $\lceil \Delta'_{I} / T_s \rceil$ of filter $\mathcal{F}_{n}$ mainly contributes to regenerate the part of the self-interference signal that arrives with delay closest to $\Delta^{(n)}_{R}$.
Consequently, for each filter $\mathcal{F}_{n}$ the tap number $\lceil \Delta'_{I} / T_s \rceil$ and neighboring taps should be given larger weights so that they are larger contributors for the regeneration of the self-interference signal arriving with delay close to $\Delta^{(n)}_{R}$.
This weighting can be performed via a multiplicative factor, which we denote as $\bm{p}^{(n)}$ that scales either the filter weights or the filter input of the reference $n$.
We have found via simulations that scaling the filter input results in faster convergence.
Following the above insight, we propose a customized NLMS algorithm as explained below.

Let $t_k = kT_s$ be the sample times for $k \in \mathbb Z$.
Consider an architecture as in Fig.~\ref{fig:multitap_bd} and assume $\Delta^{(n)}_{R}=(n-1)T'$ for $n=1, \ldots, N$; consecutive references therefore have a delay difference of $T'$.
Assume $T' \geq T_s$.
Define $\bm{y}^{(n)}_{R}(t_k)$ as the vector representing the last $L$ samples received from reference~$n$
\begin{align}
	\bm{y}^{(n)}_{R}(t_k) &= \big[ y^{(n)}_{R}(t_k),  y^{(n)}_{R}(t_k-T_s), ... , y^{(n)}_{R}(t_k-(L-1)T_s)\big] \nonumber
\end{align}
The $j$-th entry of $\bm{y}^{(n)}_{R}(t_k)$, which we denote as $[\bm{y}^{(n)}_{R}(t_k)]_{j}$, is given by
\begin{align}
	\left[\bm{y}^{(n)}_{R}(t_k)\right]_{j}=h^{(n)}_{R} e^{\jmath\phi(t_k-(j-1)T_s)}x\left(t_k-(j-1)T_s-\Delta^{(n)}_{R}\right) \nonumber
\end{align}
Let us define $\bm{u}(t_k)$ as the vector with $NL$ elements that is input to the bank of adaptive filters at time $t_k$.
Furthermore, we define $\bm{g}(t_k)$, with $NL$ elements, as the vector representing the values of the filter taps for the $N$ filters at time $t_k$.
Our proposed NLMS-like adaptive filter approach for self-interference regeneration based on multiple references is shown in Alg.~\ref{nlms}.
We use $\odot$ as the Hadamard entrywise product between matrices or vectors, and $\mathbf{vec}(\mathbf A)$ as the vectorize operation that stacks the columns of matrix $\mathbf A$ in a column vector.
We note that our proposed construction of $\bm{p}^{(n)}$ in Alg.~\ref{nlms} is based on a heuristic approach derived from our earlier observations.
Obtaining the optimal $\bm{p}^{(n)}$ is currently part of our future work.

\renewcommand{\algorithmicensure}{\textbf{Output:}}
\renewcommand{\algorithmicrequire}{\textbf{Input:}}
\begin{algorithm}[t]                      
\caption{Multitap self-interference canceller algorithm}          
\label{nlms}                           
\begin{algorithmic}                    
	\REQUIRE{Step size $\mu$.}
	\REQUIRE{Received signal $y_{I}(t_k)$.} 
	\REQUIRE{Reference signals $\bm{y}^{(n)}_{R}(t_k)$ for all $1 \leq n \leq N$.}
    \STATE {\bf{(A)} \it{Compute mask for filter input.}}
    	\STATE Initialize $[\bm{p}^{(n)}]_l=0$, for $1 \leq l \leq L$ and $1 \leq n \leq N$.
    	\STATE Compute $\delta_{1}= \lceil \frac{\Delta'_{I}}{T_s} \rceil$ and $\delta_{2}=\lceil \frac{T'}{(2T_s)}\rceil $. 
    	\FOR{$1 \leq n \leq N$}
    		\STATE {\textbf{if} $\left( T'  > 2T_s\right)$ \textbf{then} $[\bm{p}^{(n)}]_l=1$ for $ \delta_{1}-\delta_{2} \leq l \leq  \delta_{1}+\delta_{2}$}
    		\STATE {\textbf{elsif} $\left( T' = 2T_s\right)$ \textbf{then} $[\bm{p}^{(n)}]_l=1$ for $\delta_{1}  \leq l \leq \delta_{1}+1$}
    		\STATE {\textbf{else} $[\bm{p}^{(n)}]_l=1$ for $ l =\delta_{1}$} 
    	\ENDFOR
    \STATE {\bf{(B)} \it{Run adaptive algorithm.}}	
    \STATE Initialize $\bm{g}(t_k) \Leftarrow \bm 0$
    \FORALL{$k \in \mathbb N$}
    	\STATE $\bm Y(t_k) \Leftarrow \left[\bm{y}^{(1)}_{R}(t_k) \odot \bm{p}^{(1)}, \ldots, \bm{y}^{(N)}_{R}(t_k) \odot \bm{p}^{(N)}\right]$
    	\STATE $\bm{u}(t_k) \Leftarrow \mathbf{vec} \,( \bm Y(t_k)) $
    	\STATE $\widehat{y}_{I}(t_k) \Leftarrow \bm{g}^{H}(t_k) \bm{u}(t_k)$
        \STATE $e(t_k) \Leftarrow y_{I}(t_k) - \widehat{y}_{I}(t_k)$
        \STATE $\bm{g}(t_{k+1}) \Leftarrow \bm{g}(t_k)- \mu \cfrac{e(t_k)^* \bm{u}(t_k)}{\bm{u}(t_k)^H\bm{u}(t_k)}$
    \ENDFOR
    \ENSURE{Estimated self-interference $\widehat{y}_{I}(t_k)$ for all $k \in \mathbb N$.}
\end{algorithmic}
\end{algorithm}

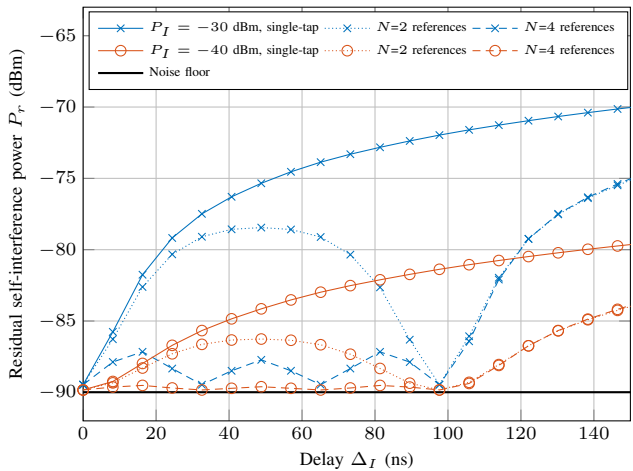
\begin{figure}[t!]
	\centering
	\scriptsize
	\begin{tikzpicture}
	\begin{axis}[
		width=\figurewidth,
		height=0.8\figurewidth,
		xmajorgrids,
		ymajorgrids,
		xlabel=Delay $\Delta_I$ (ns),
		ylabel=Residual self-interference power $P_r$ (dBm),
		legend entries={
			{$P_I = -30$ dBm, single-tap},
			{$N$=$2$ references},
			{$N$=$4$ references},
			{$P_I = -40$ dBm, single-tap},
			{$N$=$2$ references},
			{$N$=$4$ references},
			{Noise floor},
		},
		legend columns=3,
		mark options={solid},
		xmin=0,
		xmax=150,
		ymin=-92,
		ymax=-63,
		legend style={at={(0.01, 0.99)}, anchor=north west, font=\tiny},
		legend cell align=left,
	]
		\addplot+[color=col1, solid, mark=x] table[x index=0, y index=1] {spawc_mt_sp.data};
		\addplot+[color=col1, densely dotted, mark=x] table[x index=0, y index=2] {spawc_mt_sp.data};
		\addplot+[color=col1, densely dashed, mark=x] table[x index=0, y index=3] {spawc_mt_sp.data};
		\addplot+[color=col2, solid, mark=o] table[x index=0, y index=4] {spawc_mt_sp.data};
		\addplot+[color=col2, densely dotted, mark=o] table[x index=0, y index=5] {spawc_mt_sp.data};
		\addplot+[color=col2, densely dashed, mark=o] table[x index=0, y index=6] {spawc_mt_sp.data};
		\addplot+[color=black, solid, thick, mark=none] table[x index=0, y index=7] {spawc_mt_sp.data};
	\end{axis}
	\end{tikzpicture}
	\caption{Comparison of the residual interference power achieved by the single-tap canceller versus the residual achieved by the proposed multi-tap canceller for the scenario of a single path channel with a varying path delay $\Delta_I$. The $N$ reference tap delays are fixed and spaced equally between 0 and 100 ns.}
	\label{fig:multitap_singlepath}
\end{figure}

\begin{figure}[b!]
	\centering
	\scriptsize
	\begin{tikzpicture}
	\begin{axis}[
		width=\figurewidth,
		height=0.7\figurewidth,
		xmajorgrids,
		ymajorgrids,
		xlabel=Delay spread $|\Delta^{(2)}_{I} - \Delta^{(1)}_{I}|$ (ns),
		ylabel=Residual self-interference power $P_r$ (dBm),
		legend entries={
			{single-tap},
			{$N = 2$ references},
			{$N = 4$ references},
			{No phase noise},
			{Noise floor},
		},
		mark options={solid},
		xmin=0,
		xmax=150,
		ymin=-92,
		ymax=-73,
		legend style={at={(0.01, 0.99)}, anchor=north west, font=\tiny},
		legend cell align=left,
	]
		\addplot+[color=col1, solid, mark=x] table[x index=0, y index=1] {spawc_mt_mp.data};
		\addplot+[color=col1, densely dotted, mark=x] table[x index=0, y index=2] {spawc_mt_mp.data};
		\addplot+[color=col1, densely dashed, mark=x] table[x index=0, y index=3] {spawc_mt_mp.data};
		\addplot+[color=col2, solid] table[x index=0, y index=4] {spawc_mt_mp.data};
		\addplot+[color=black, solid, thick, mark=none] table[x index=0, y index=5] {spawc_mt_mp.data};
	\end{axis}
	\end{tikzpicture}
	\caption{ Comparison of the residual interference power achieved by the single-tap canceller versus the residual achieved by the proposed multi-tap canceller for the scenario of a 2-path channel with absolute delay spread $|\Delta^{(2)}_{I} - \Delta^{(1)}_{I}|$.  The $N$ reference tap delays are fixed and spaced equally between 0 and 100~ns.
	}
	\label{fig:multitap_multipath}
\end{figure}
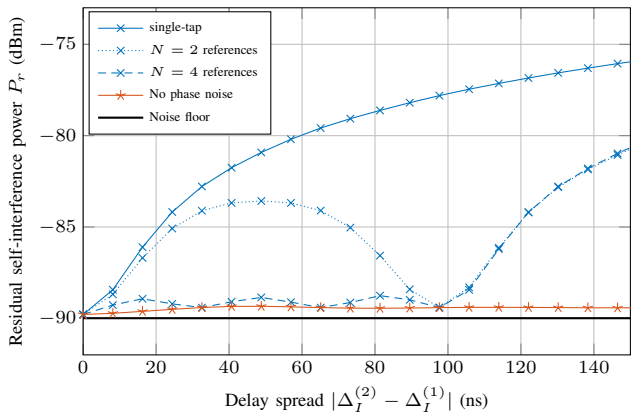

Simulation results shown in Fig.~\ref{fig:multitap_singlepath} compare the baseline single-tap canceller explained in Section~\ref{sec:pn_effect} versus our proposed multi-tap canceller for a single path self-interference channel.
The results demonstrate the improvement thanks to our proposed multi-tap approach.
We observe that when the delay of the self-interference channel $\Delta_{I}$ matches one of the reference delays $\Delta^{(n)}_{R}$, the residual interference falls back to the noise floor.
Fig.~\ref{fig:multitap_multipath} shows results for a channel with two paths as was assumed and described for the simulations of Fig.~\ref{fig:singletap_multipath}.
Notice the performance improvement thanks to the use of multiple references.
With four taps, our design is almost reducing the self-interference to the noise floor for delay spreads up to 100~ns---the delay of the last reference.
The performance degrades beyond this value.

Finally, we point out that all the filters evaluated in our simulations have a total of 32 taps. For our design in Alg.~\ref{nlms} the vector $\bm{g}(t_k)$ has  $NL$ elements.
However, we can simplify the filter and reduce the total equalizer length to only $L$ elements since the mask $\bm{p}$ on the input signal has many zero values.
The details of this simplification are not included due to lack of space, and will be detailed in an upcoming work.

%% file: conclusion.tex
\section{Conclusion}
In this paper we have presented a new architecture for digital self-interference cancellation that is based on a novel use of multiple reference signals and a new customized LMS filtering for self-interference regeneration. This architecture achieves larger levels of cancellation than state-of-the-art single-tap solutions which, as we have demonstrated, can only match the delay of a single path channel hence their performance degrades in the presence of phase noise for multi path channels with non zero delay spread.

%% file: appendix.tex

\appendix
Our simulations correspond to an OFDM signal with 20MHz bandwidth.
We use an FFT size of 2048, sampling frequency of $122.88$~MHz for the A/D and D/A converters and a sampling frequency of $30.72$~MHz before (after) upsampling at the transmitter (downsampling at the receiver) which corresponds to an upsampling (downsampling) factor of 4.
We model independent phase noise processes at the transmitter and receiver and consider the MAX2829 oscillator used in the WARP boards, which have been used in several demonstrations of full-duplex systems \cite{Sahai2013,Duarte2012,Bharadia2013}.
For simplicity, we model the oscillator as a Wiener process; in this case the difference process $\phi(t) - \phi(t-\Delta)$ is distributed as $\mathcal N(0, \sigma^2_{\epsilon}|\Delta|)$.
We compute $\sigma^2_{\epsilon}$ from the PSD provided in the datasheet \cite{Oscillator} and its value is set to 8.5095e-5, in units of squared radians.
In our simulations, the phase noise is incorporated at sampling frequency of $122.88$~MHz and the self-interference regeneration and cancellation occurs on the signals running at $30.72$~MHz.
The noise floor is equal to $-90$~dBm in all simulations.